\documentclass[cits]{pos}

\title{
\rule{0cm}{2.5cm}\vspace{-4.5cm}\\{ 
\hfill \it \normalsize SB/F/378-10
 } \vspace{2cm}\\Jet and $W/Z$ Production at Hadron Colliders}

\ShortTitle{Jet and $W/Z$ Production}

\author{\speaker{Fernando Febres Cordero}${}^{\ a,b}$\\
        \llap{$^a$}Universidad Sim\'on Bol\'{\i}var, Departamento de F\'{\i}sica, Apartado 89000, Caracas 1080A, Venezuela\\
        \llap{$^b$}Department of Physics and Astronomy, UCLA, Los Angeles, CA 90095-1547, USA\\
        E-mail: \email{ffebres@usb.ve}}

\abstract{The start of the physics program at the LHC has added great impetus in the development of
powerful theoretical tools to meet the many challenges that this collider brings. The production of
jets and weak vector bosons is at the center of most analyses, from machine performance
to new physics searches. In this talk we review some recent advances in the study of jets, in the
computation of quantum corrections to processes with large jet multiplicity and their impact in
$W/Z+{\rm jets}$ and $W/Z+b-{\rm jets}$ production at the Tevatron and the LHC.}

\FullConference{XXth Hadron Collider Physics Symposium \\
		 November 16 -- 20, 2009 \\
		 Evian, France}

\begin{document}

\section{Introduction}
Our understanding of the behavior of fundamental particles will be newly tested with the start of
the LHC. With every step into higher energies we will be able to keep exploring the validity of the
Standard Model (SM), and especially of its mechanism for Electroweak Symmetry Breaking, as well as
testing possible scenarios of physics beyond the SM (BSM). In these tasks studying the production of weak
vector bosons and jets is fundamental, given that they are our basic tools to extract information
from hard interactions. Understanding Drell-Yan (DY) production is for example highly beneficial because
of the close connection that these processes have to the determination of important quantities like
luminosity and parton distribution functions.

As the strong coupling constant is relatively large at the scales of interest, the inclusion of
perturbative QCD corrections to differential cross sections is necessary in order to successfully
describe hadron collider data (see e.g. Ref.~\cite{Aaltonen:2007cp,Aaltonen:2007ip}). For instance,
it is well known that DY processes receive large next-to-leading order (NLO) QCD corrections,
in some cases up to factors of two or more. In part this is also understood due to the strong kinematical
constraints for leading order (LO) production and to the opening of new production channels at NLO.
Nevertheless, one then has to show the validity of the perturbative
expansion. Currently next-to-next-to-leading order (NNLO) QCD corrections to DY processes are
available, and at this level it is shown that the perturbative expansion stabilizes. Programs like
{\tt Vrap}~\cite{Anastasiou:2003ds} and {\tt FEWZ}~\cite{Melnikov:2006kv} (see also
Ref.~\cite{Catani:2009sm}) compute related observables at NNLO precision. 

Having tools that would allow for computations of NLO QCD corrections for a large variety of
processes of interest is then highly desirable, especially for processes with large jet multiplicity
which suffer from large uncertainties in normalization and shape of
distributions~\cite{Bern:2008ef}.

In the following we will review central topics that have seen considerable progress over the last
years to improve our theoretical control over the production of jets and weak vector bosons, in particular
as the number of jets increases.

\section{IR Safe Jet Algorithms (Fast!)}
Except for a handful of examples, all studies including jets that have been performed by the
Tevatron experiments, CDF and D0, have used cone like jet algorithms which are known to suffer from
infrared unsafety. From a theoretical point of view, the use of an infrared unsafe algorithm in a
perturbative computation spoils the order by order cancellation of infrared (soft and/or collinear)
divergences. This turns even LO computations meaningless, for sufficiently large multiplicities, as
the effective expansion parameter of the perturbative series becomes of $O(1)$ (for a nice review,
see Ref.~\cite{Salam:2009jx}). But technical constraints kept the experimental collaborations from
using existing IR safe algorithms, like $k_T$ or seedless cone jet algorithms. Basically the
computational need for clustering of those algorithms grew too quickly with the number of input
towers (or particles) $N$: past implementations of the $k_T$ and seedless cone jet algorithms had a
$N^3$ and exponential time scaling respectively.  But with the help of sequential recombination
algorithms and computational geometry techniques, $N\ln(N)$ implementations of the $k_T$ algorithm
(and of related sequential jet algorithms)~\cite{Cacciari:2005hq,Cacciari:2008gp}, as well as a
$N^2\ln(N)$ implementation of a seedless cone algorithm (SISCone)~\cite{Salam:2007xv} became
available.
\begin{figure}[ht]
\begin{center}
\includegraphics[scale=0.4]{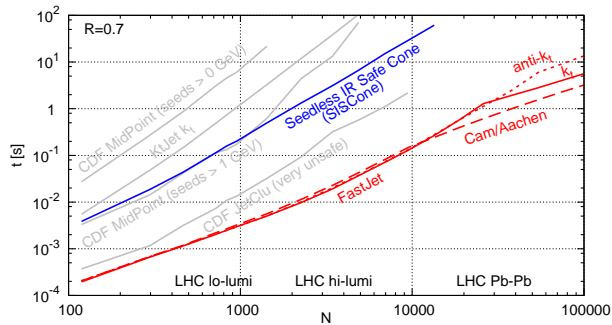}
\caption{Clustering time as a function of the number of input momenta for several jet algorithms. IR
safe algorithms are shown in dark colors (read and blue), and IR unsafe algorithms in light grey.
Taken from Ref.~\cite{Salam:2009jx}.}
\label{ja_time_sc}
\end{center}
\end{figure}

As Fig.~\ref{ja_time_sc} shows, new implementations of IR safe jet algorithms perform similarly or
better than commonly used IR unsafe cone algorithms. These then allow the ATLAS and CMS
collaborations to use IR safe jet algorithms, even as their default jet algorithm. Indeed, a large
amount of fast IR safe jet algorithms are now on the market, and with them a lot of new ideas have
appeared (like pruning, filtering, variable-R algorithms, etc.) that should allow for optimizations
in jet definitions for specific studies (see for example the review Ref.~\cite{Salam:2009jx}).

\section{NLO QCD corrections to $W/Z+n\ {\rm jets}$ ($n=1,2,3$) at Hadron Colliders}
In 2009 the first full NLO QCD corrections to hadron collider processes
with four particles in the final state became available, including $t\bar{t}b\bar{b}$ production~\cite{ttbbNLO},
$W+3$ jets production~\cite{WjetsatNLO,WjetslcNLO} and $Z+3$ jets production~\cite{Berger:2009dq}.
A good part of the progress has been due to the use of new on-shell techniques (for a recent review
see Ref.~\cite{Berger:2009zb}).

\begin{figure}[ht]
\begin{center}
\includegraphics[clip,scale=0.3]{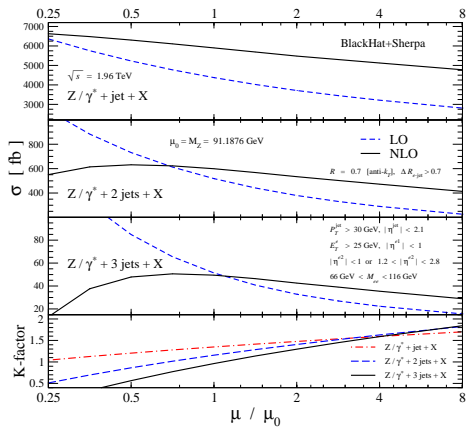}
\caption{Total cross section dependence on renormalization and factorization scales
($\mu_r=\mu_f=\mu$) at LO (dashed-blue) and NLO (solid-black) for $Z/\gamma^{\ \!*}+n$ jets ($n=1,2,3$)
production at the Tevatron. Bottom panel shows K-factors for each jet multiplicity. Taken from
Ref.~\cite{Berger:2009dq}.}
\label{Zjets_Tev_sd}
\end{center}
\end{figure}

Understanding $W/Z+$ jet production is especially important given that it leads to signals of missing
energy plus jets, in itself a typical signature associated to physics BSM (see for example
Refs.~\cite{BSMrev}). To disentangle signal and backgrounds, NLO predictions are needed especially for
high jet multiplicity processes, given the large theoretical uncertainty associated to LO based
predictions.  This is explicitly shown in Fig.~\ref{Zjets_Tev_sd}, where we see how the dependence
on factorization and renormalization scales of the K-factor for the total cross sections for
$Z/\gamma^{\ \!*}+n$ jets ($n=1,2,3$) production at the Tevatron increases with the numbers of jets.

\begin{figure}[ht]
\begin{center}
\includegraphics[clip,scale=0.3]{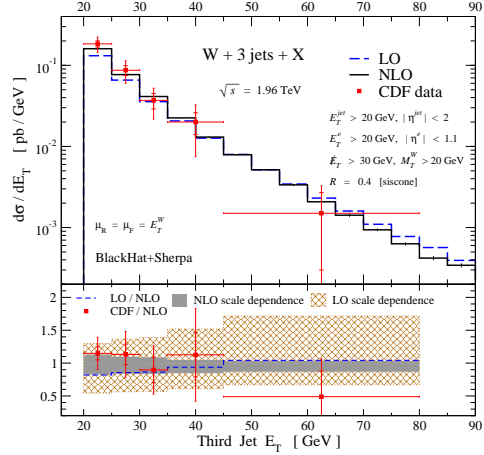}
\caption{NLO QCD corrections to third jet $E_T$ distributions in $W+jets$ production compared to CDF
data~\cite{Aaltonen:2007ip}. Taken from Ref.~\cite{WjetsatNLO}.}
\label{W3j_with_CDF_data}
\end{center}
\end{figure}

\begin{figure}[ht]
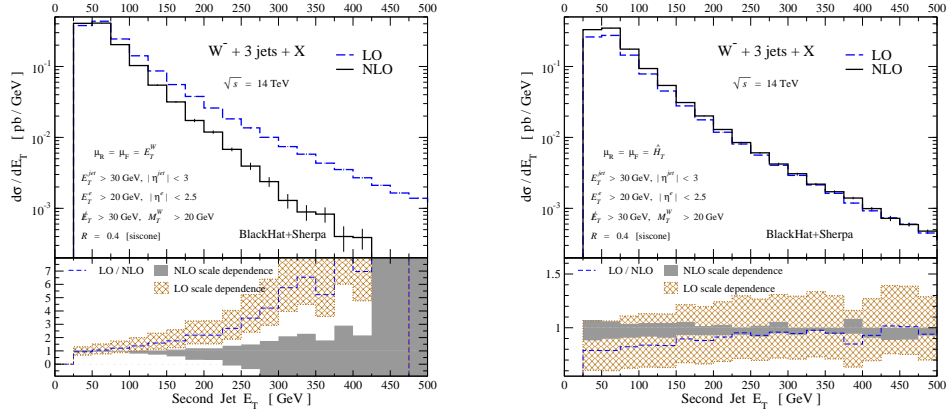

\begin{center}
\includegraphics[clip,scale=0.27]{Wm3jLHC_ETWmu_siscone_eb_jets_jet_1_1_Et_2.eps}
\hspace{1cm}\includegraphics[clip,scale=0.27]{Wm3jLHC_HTmu_siscone_eb_jets_jet_1_1_Et_2.eps}
\caption{Second Jet $E_T$ distribution for $W+3$ jets production at the LHC, computed with a
dynamical scale set to $E_T^W$ (left) and ${\hat H}_T$ (right). The latter clearly gives more stable
results both at LO and NLO. Taken from Ref.~\cite{WjetsatNLO}.}
\label{Wm3j_LHC_scales}
\end{center}
\end{figure}

NLO corrections give then the first quantitatively reliable prediction of total rates, and even
more, their reduced sensibility to unphysical scales implies improved predictions for the shape of
distributions as well. 
In Fig.~\ref{W3j_with_CDF_data} we see for example how NLO predictions fit well CDF's
data~\cite{Aaltonen:2007ip} for the $E_T$ distribution of the third jet in inclusive $W+$ jets
production. In part the LO
difference in shape with CDF's data, is mostly due to a poor choice of dynamical scale, namely $E_T^W$.
This choice can be shown, armed with NLO predictions, to be even worse for the LHC, as it will
sample a larger dynamical range. For example, the left panel of Fig.~\ref{Wm3j_LHC_scales} shows a
large shape change from LO to NLO, and more troublesome it shows the NLO prediction turning
negative at large second jet $E_T$! This as a consequence of having introduced large logs in the
computation, due to poor choice of dynamical scales~\cite{WjetsatNLO}. In the right panel we show
similar results with the dynamical scale set to ${\hat H}_T$, which behave much better. Indeed such
choice leads to fairly flat bin-by-bin K-factors over full phase space (a nice feature when NLO effects are
introduced in Monte Carlo programs via a global K-factor rescaling).

\begin{figure}[ht]
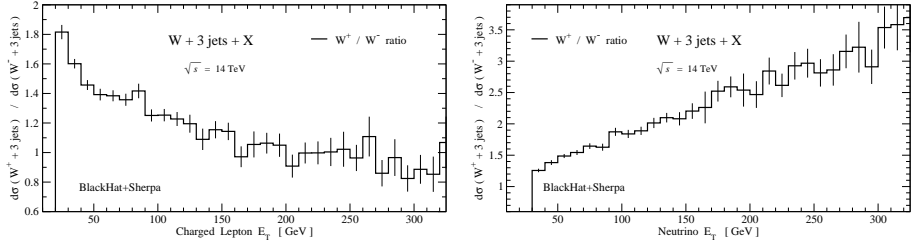

\begin{center}
\includegraphics[clip,scale=0.25]{Wp3jLHCHT_over_Wm3jLHCHT_central_mu_siscone_PTchargedlepton.eps}
\hspace{0.2cm}\includegraphics[clip,scale=0.25]{Wp3jLHCHT_over_Wm3jLHCHT_central_mu_siscone_PTneutrino.eps}
\caption{Charged lepton (left) and neutrino (right) $E_T$ distributions for the ratio ($W^++3$
jets)/($W^-+3$ jets) at the LHC. Large asymmetries are found due to left-handed polarization of the parents
$W^+$ and $W^-$. Taken from Ref.~\cite{WjetsatNLO}.}
\label{Wp3j_LHC_polarization}
\end{center}
\end{figure}

To close this section, we mention an interesting feature found for the QCD production of $W+$
jets~\cite{WjetsatNLO}. At the LHC the production of $W+$ jets shows both $W^+$ and $W^-$ being
produced with left-handed polarization for large $P_T^W$. This effect is fairly independent of the
number of jets considered, and is found both at LO and NLO. In Fig.~\ref{Wp3j_LHC_polarization} we
show how this effect results in an asymmetry in the $E_T$ distributions of the decay leptons in
$W^+$ and $W^-$ production.  We notice that this feature does not appear for example when the
leptons and jets come from top production or other BSM signals, where $W^+$ tends to be left-handed
while $W^-$ tends to be right-handed for large $P_T^W$, and giving then no asymmetries as the ones
in Fig.~\ref{Wp3j_LHC_polarization}.

\section{$W$ Associate Production to $b$-jets}
A recent measurement of the cross section of $W$ boson production in association with one or two
$b$-jets by the CDF collaboration at the Tevatron finds~\cite{Aaltonen:2009qi}
\begin{equation}
\sigma_{b-\mathrm{jets}}\times \mathcal{B} (W\rightarrow \ell \nu ) (\mathrm{CDF})
= 2.74 \pm 0.27\textrm{(stat.)} \pm 0.42\textrm{(syst.)}~\mathrm{pb}\ .
\end{equation}
This $b$-jet cross section includes $Wb\bar b$ and $W+1 b$-jet contributions, and the NLO QCD
predictions for both signatures have to be considered. The NLO QCD prediction for $Wb\bar b$
production is based on Refs.~\cite{WbbNLO} and the one for $W+1 b$-jet
production on Ref.~\cite{Campbell:2008hh}, where in both cases events with a non-$b$-jet that
result in a three-jet event are discarded. 
Combining the results of predictions for $Wb\bar b$ and $W+1 b$-jet production,
yields the
following NLO QCD predictions (with $\mu_r=\mu_f=M_W$)~\cite{Campbell:2008hh}:
\begin{equation}
\sigma_{b-\mathrm{jets}}\times \mathcal{B} (W\rightarrow \ell \nu ) (\mathrm{NLO\; QCD})
= 1.22 \pm 0.14~\mathrm{pb} \; .
\end{equation}
Together with the LO prediction of $0.91^{+0.29}_{-0.20}$~pb (including scale uncertainties) this
results in a moderate K-factor of about 1.35. There is then a clear discrepancy between theory and
experiment, even when comparing to shower montecarlo results~\cite{Aaltonen:2009qi}. The origin of
the discrepancy is still an open problem~\footnote{See talks given at the workshop {\em
Northwest Terascale Research Projects $W + b$ quark physics at the LHC}, held at the University of
Oregon, http://physics.uoregon.edu/\textasciitilde soper/TeraWWW2.}.

\section*{Acknowledgments}
We would like to thank the organizers of the HCP2009 Symposium for their kind invitation and
their work to prepare such wonderful conference. We thank too C.F. Berger, Z. Bern, J. Campbell, L.
Dixon, K. Ellis, D. Forde, T.  Gleisberg, H. Ita, D. Kosower, D. Ma\^{\i}tre, F. Maltoni, L. Reina,
D. Wackeroth and S. Willenbrock for collaboration in several of the topics presented in this talk.

\end{document}